\documentclass[runningheads]{llncs}
\usepackage{amsmath}
\usepackage{amssymb}
\usepackage{algorithmic}
\usepackage{algorithm}
\usepackage{graphicx}
\usepackage{xcolor}
\usepackage{algorithmic}
\usepackage{algorithm}
\usepackage{multicol}  
\usepackage{multirow}
\usepackage{diagbox} 
\usepackage{booktabs} 
\usepackage{mathrsfs}
\usepackage[backref]{hyperref} 
\usepackage{tikz}
\usepackage{makecell}
\usepackage{multirow}
\usetikzlibrary{arrows, automata}

\newtheorem{theorem*}{Theorem}
\newtheorem{observation}{Observation}

\renewcommand{\epsilon}{\varepsilon}

\newcommand{\n}{\mathbb{N}}
\newcommand{\reals}{\mathbb{R}}
\renewcommand{\r}{\mathbb{R}}
\newcommand{\z}{\mathbb{Z}}

\newcommand{\iso}{\textrm{Iso}}
\newcommand{\Exp}{\textrm{Exp}}

\newcommand{\ds}{\ensuremath{\mathcal{D(S)}}}

\renewcommand{\tt}{\ensuremath{\mathscr{T}}}
\newcommand{\ii}{\ensuremath{\mathscr{I}}}

\newcommand{\ap}{AP}

\renewcommand{\c}{\ensuremath{\mathbb{C}}}
\renewcommand{\t}{\ensuremath{\mathcal{T}}}

\renewcommand{\a}{\ensuremath{\mathbb{A}}}

\renewcommand{\nu}{\ensuremath{\textrm{nu}}}

\newcommand{\f}{\ensuremath{\mathbb{F}}}

\newcommand{\m}{\ensuremath{\mathcal{M}}}
\newcommand{\sm}{\ensuremath{\mathcal{SM}}}

\newcommand{\q}{\ensuremath{\mathbb{Q}}}
\renewcommand{\i}{\ensuremath{\mathcal{I}}}

\newcommand{\s}{\ensuremath{\mathbb{S}}}

\newcommand{\vp}{\varphi}
\newcommand{\pair}[2]{\langle #1 , #2 \rangle}

\newcommand{\todo}[1]{\color{red} {TODO: #1 :TODO} \color{black} }

\begin{document}
\title{A Probabilistic Logic for Verifying  Continuous-time Markov Chains}  
\author{
Ji Guan\inst{1}
\and
Nengkun Yu \inst{2}}
\authorrunning{J. Guan and N. Yu}
\institute{State Key Laboratory of Computer Science, Institute of Software,
Chinese Academy of Sciences, Beijing,
China.\\
guanji1992@gmail.com
 \and Centre for Quantum Software and Information,
 University of Technology Sydney, Sydney, Australia.
 \\
nengkunyu@gmail.com}
% \email{nengkunyu@gmail.com}
%}
%\email{guanji1992@gmail.com}
%\date{\today}
\maketitle
\begin{abstract}
A continuous-time Markov chain (CTMC) execution is a continuous class of probability distributions over states. This paper proposes a probabilistic linear-time temporal logic, namely continuous-time linear logic (CLL), to reason about the probability distribution execution of CTMCs.  We define the syntax of CLL on the space of probability distributions. The syntax of CLL includes multiphase timed until formulas, and the semantics of CLL allows time reset to study relatively temporal properties. We derive a corresponding model-checking algorithm for CLL formulas. The correctness of the model-checking algorithm depends on Schanuel's conjecture, a central open problem in transcendental number theory. Furthermore, we provide a running example of CTMCs to illustrate our method.
\end{abstract}

\section{Introduction}
As a popular model of probabilistic continuous-time systems, \emph{continuous-time Markov chains (CTMCs)} have been extensively studied since Kolmogorov~\cite{kolmogoroff1931analytischen}.
In the recent 20 years, probabilistic continuous-time model checking receives much attention. Adopting \emph{probabilistic computational tree logic (PCTL)}~\cite{hansson1994logic} to this context with extra \emph{multiphase timed until formulas} $\Phi_1 U^{\t_1} \Phi_2 \cdots U^{\t_K} \Phi_{K+1}$,
for state formula $\Phi$ and time interval $\t$, Aziz et al. proposed \emph{continuous stochastic logic (CSL)} to specify the branching-time properties of CTMCs and the model-checking problem for CSL is decidable~\cite{aziz2000model}. After that, efficient model-checking algorithms were developed by transient analysis of CTMCs using uniformization~\cite{baier2003model} and stratification~\cite{zhang2011automata} for a restricted version (path formulas are restricted to single until formulas $\Phi_1 U^{\i} \Phi_2$) and a full version of CSL, respectively.
These algorithms have been practically implemented in model checkers PRISM~\cite{kwiatkowska2002prism}, MRMC~\cite{katoen2011ins} and STORM \cite{dehnert2017storm}.
Further details can be found in an excellent survey~\cite{katoen2016probabilistic}.

There are also different ways to specify the linear-time properties of CTMCs. Timed automata were first used to achieve this task~\cite{barbot2011efficient,chen2009quantitative,chen2011model,chen2011observing,feng2018monitoring}, and then \emph{metric temporal logic (MTL)}~\cite{chen2011time} was also considered in this context. Subsequently, the probability of ``the system being in \emph{state $s_0$} within five-time units after having continuously remained in \emph{state $s_1$}" can be computed. However, some statements cannot be specified and verified because of the lack of a probabilistic linear-time temporal logic, for instance
``the system being in \emph{state $s_0$} with high probability ($\geq 0.9$) within five-time units after having continuously remained in \emph{state $s_1$} with low probability ($\leq 0.1$)". Furthermore, this probabilistic property cannot be expressed by CSL because CSL cannot express properties that are defined across several state transitions of the same time length in the execution of a CTMC.

In this paper, targeting to express the mentioned probabilistic linear-time properties,
we introduce \emph{continuous-time linear logic (CLL)}. In particular, we adopt the viewpoint used in ~\cite{agrawal2015approximate} by regarding CTMCs as transformers of probability distributions over states. CLL studies the properties of the probability distribution execution generated by a given initial probability distribution over time. By the fundamental difference between the views of state executions and probability distribution executions of CTMCs, CLL and CSL are incomparable and complementary, as the relation between \emph{probabilistic linear-time temporal logic (PLTL)} and PCTL in model checking discrete-time Markov chains~\cite[Section 3.3]{agrawal2015approximate}.

The atomic propositions of CLL are explained on the space of probability distributions over states of CTMCs. We apply the method of symbolic dynamics to the probability distributions of CTMCs. To be specific, we symbolize the probability value space $[0,1]$ into a finite set of intervals $\ii=\{\i_k\subseteq [0,1]\}_{k=1}^m$. A probability distribution $\mu$ over its set of states $S=\{s_0,s_2,\ldots,s_{d-1}\}$ is then represented symbolically as a set of symbols $$\s(\mu)=\{\pair{s}{\i}\in S\times \ii:\mu(s)\in\i\}$$
where each symbol $\pair{s}{\i}$ asserts $\mu(s)\in\i$, i.e., the probability of state $s$ in distribution $\mu$ falls in interval $\i$. For example, $\pair{s_0}{[0.9,1]}$ means the system is in \emph{state $s_0$} with a probability in $0.9$ to $1$. The symbolization idea of distributions has been considered in \cite{agrawal2015approximate}: choosing a disjoint cover of $[0,1]$: $$\mathcal{I} = \{[0,p_1), [p_1, p_2),...,[p_n,1]\}.$$ Here, we remove this restriction and enrich the expressiveness of $\ii$. A crucial fact about this symbolization is that the set $S\times \ii$ is finite. Consequently, the (probability distribution) execution path generated by an initial probability distribution $\mu$ induces a sequence of symbols in $S\times \ii$ over time. Therefore, the dynamics of CTMCs can be studied in terms of a (real-time) language over the alphabet $S\times \ii$, which is the set of atomic propositions of CLL.

Different from non-probabilistic  linear-time  temporal logics --- \emph{linear-time temporal logic (LTL)} and MTL, CLL has two types of formulas: state formulas and path formulas. The state formulas are constructed using propositional connectives. The path formulas are obtained by propositional connectives and a temporal modal operator \emph{timed until} $U^{\t}$ for a bounded time interval $\t$, as in MTL and CSL. The standard next-step temporal operator in LTL is meaningless in continuous-time systems since the time domain (real numbers) is uncountable. As a result, CLL can express the above mentioned probabilistic property ``the system is at \emph{state $s_0$} with high probability ($\geq 0.9$)  within 5 time units after having continuously remained at \emph{state $s_1$} with low probability ($\leq 0.1$)" in a path formula: $$\vp=\pair{s_1}{[0,0.1]}U^{[0,5]}\pair{s_0}{[0.9,1]}.$$
In this single until formula, there is a time instant $0\leq t\leq 5$ at which \emph{state} $s_1$ with low probability  transits to \emph{state} $s_0$ with high probability. Then we illustrate this on the following timeline.
\[
\begin{tikzpicture}
\draw (-0.5,0)--(0,0)--(1,0)--(2,0)--(3,0)--(4,0)--(6,0)--(10.5,0); 
\fill (-0.5,0) circle (1pt);
\fill (2.5,0) circle (1pt);
%\fill (3,0) circle (1pt);
\node[rotate = 0] at (1, -0.5) {$\underbrace{\hspace{3cm}}_{\pair{s_1}{[0,0.1]}}$};
\node[rotate = 0] at (2.95, 0.3) {$\downarrow{t\leq 5}$};
\node[rotate = 0] at (-0.35, 0.3) {$\downarrow 0$ };
\node[rotate = 0] at (3.4, -0.3) {$\uparrow \pair{s_0}{[0.9,1]}$ };
\end{tikzpicture}
\]
Furthermore, CLL allows \emph{multiphase timed until formulas}.
The semantics of the formulas 
focuses on relative time intervals, i.e., time can be reset as in timed automata~\cite{Alur94atheory,alur1993parametric}, while those of CSL \cite{aziz2000model} are for absolute time intervals. Subsequently, CLL can express not only \emph{relatively} but also \emph{absolutely} temporal properties of CTMCs. 

We illustrate the significant difference between \emph{relatively}  temporal properties and \emph{absolutely} temporal properties of CTMCs.
For instance, ``before probability distributions transition $\vp$ happening in 3 to 7 time units, the system always stays at \emph{state $s_0$} with  a high probability $(\geq0.9)$" can be formalized  in path formulae  \[\vp'=\pair{s_0}{[0.9,1]}U^{[3,7]}(\pair{s_1}{[0,0.1]}U^{[0,5]}\pair{s_0}{[0.9,1]}).\]
As we can see, there are two time instants, namely $t_1$ and $t_2$, happening distribution transitions. Time is reset to 0 after the first distribution transition happens  and thus $t_2$ is relative to $t_1$. More clearly, we depict this on the following timeline.
\[
\begin{tikzpicture}
\draw (-5,0)--(0,0)--(1,0)--(2,0)--(3,0)--(4,0)--(6,0); 
\fill (-0.5,0) circle (1pt);
\fill (2.5,0) circle (1pt);
%\fill (3,0) circle (1pt);
\fill (-2.5,0) circle (1pt);
\fill (-5,0) circle (1pt);
\node[rotate = 0] at (-1.5, -0.5) {$\underbrace{\hspace{2cm}}_{\pair{s_0}{[0.9,1]}}$};
\node[rotate = 0] at (1, -0.5) {$\underbrace{\hspace{3cm}}_{\pair{s_1}{[0,0.1]}}$};
\node[rotate = 0] at (3.6, 0.3) {$\downarrow{(t_2+t_1)\leq  12}$};
\node[rotate = 0] at (-3.75, 0.5) {$\overbrace{\hspace{2.5cm}}^{= 3}$};
\node[rotate = 0] at (-4.8, -0.3) {$\uparrow$ 0s};
\node[rotate = 0] at (0.05, 0.3) {$\downarrow t_1\leq 7$ };
\node[rotate = 0] at (3.4, -0.3) {$\uparrow \pair{s_0}{[0.9,1]}$ };
\end{tikzpicture}
\]
An absolute version is ``probability distribution transition $\vp$ happens and the system always stays at \emph{state} $s_0$ with a high probability ($\geq$ 0.9) in 3 to 7 time units'' 
\[\vp''=\Box^{[3,7]}\pair{s_0}{[0.9,1]}\land \pair{s_1}{[0,0.1]}U^{[0,5]}\pair{s_0}{[0.9,1]}).\]
We can get a clear timeline representation by simply adding $\Box^{[3,7]}\pair{s_0}{[0.9,1]}$ to that of $\vp$. Assume that $t<3$,
\[
\begin{tikzpicture}
\draw (-0.5,0)--(0,0)--(1,0)--(2,0)--(3,0)--(4,0)--(6,0)--(10.5,0); 
\fill (-0.5,0) circle (1pt);
\fill (2.5,0) circle (1pt);
\fill (4.5,0) circle (1pt);
\fill (10,0) circle (1pt);
\node[rotate = 0] at (1, -0.5) {$\underbrace{\hspace{3cm}}_{\pair{s_1}{[0,0.1]}}$};
\node[rotate = 0] at (2.95, 0.3) {$\downarrow{t< 3}$};
\node[rotate = 0] at (-0.35, 0.3) {$\downarrow 0$ };
\node[rotate = 0] at (3.4, -0.3) {$\uparrow \pair{s_0}{[0.9,1]}$ };
\node[rotate = 0] at (4.65, 0.3) {$\downarrow 3$ };
\node[rotate = 0] at (10.15, 0.3) {$\downarrow 7$ };
\node[rotate = 0] at (7.25, -0.5) {$\underbrace{\hspace{5.5cm}}_{\pair{s_0}{[0.9,1]}}$};
\end{tikzpicture}
\]

Time reset enriches the expressiveness of CLL but introduces more difficulties to model checking CLL than CSL. We cross this by translating relative time to the absolute one. As a result,
we develop an algorithm to model check CTMCs against CLL formulas. More precisely, we reduce the model-checking problem to a reachability problem of absolute time intervals. The reachability problem corresponds to the real root isolation problem of \emph{real polynomial-exponential functions (PEFs)} over the field of algebraic numbers, an extensively studied question in recent symbolic and algebraic computation community (e.g. \cite{achatz2008deciding,gan2017reachability,li2016positive}). By developing a state-of-the-art real root isolation algorithm, we resolve the latter problem under the assumption of the validity of Schanuel's conjecture, a central open question in transcendental number theory~\cite{lang1966introduction}. This conjecture has also been the footstone of the correctness of many recent model-checking algorithms, including the decidability of continuous-time Markov decision processes~\cite{majumdar2020decidability}, the synthesizing inductive invariants for continuous linear dynamical systems~\cite{almagor2020invariants}, the termination analysis for probabilistic programs with delays~\cite{xu2020time}, and reachability analysis for dynamical systems~\cite{gan2017reachability}.

In summary, the main contributions of this paper are as follows.
\begin{itemize}
	\item \emph{Introducing a probabilistic logic}, namely continuous-time linear logic (CLL), for reasoning about CTMCs;
    \item \emph{Developing} a state-of-the-art real root
isolation algorithm for PEFs over the field of algebraic numbers for checking atomic propositions of CLL;
	\item \emph{Proving} that model checking CTMCs against CLL formulas is decidable subject to Schanuel's conjecture.
\end{itemize}

\textbf{Organization of this paper.} In the next section, we give the mathematical preliminaries used in this paper. In Section~\ref{Sec:Symbolic}, we recall the view of CTMCs as distribution transformers. After that,
the symbolic dynamics of CTMCs are introduced by symbolizing distributions over states of CTMCs in Section~\ref{sec:symbolic_dynamic}. In the subsequent section, we present our continuous-time probabilistic temporal logic CLL. In Section~\ref{Sec:CLLmodelchecking}, we develop an algorithm to solve the CLL model checking problem. A case study and related works are shown in Sections~\ref{sec:case-study} and \ref{Sec:related_works}, respectively. We summarize our results and point out future research directions in the final section.

\section{Preliminaries}\label{Sec:Pre}
For the convenience of the readers, we review basic definitions and notations of number theory, particularly
Schanuel's conjecture.

Throughout this paper, we write $\c, \r, \q$ and $\a$ for the fields of all complex, real, rational and algebraic numbers, respectively. In addition, $\z$ denotes the set of all integer numbers. For $\f\in\{\c,\r,\q,\z,\a\}$, we use $\f[t]$ and $\f^{n\times m}$ to denote the set of polynomials in $t$ with coefficients in $\f$ and $n$-by-$m$ matrices with every entry in $\f$, respectively. Furthermore, for $\f\in\{\r,\q,\z\}$, we use $\f^+$ to denote the set of positive elements (including 0) of $\f$.

A bounded (time) \emph{interval} $\t$ is a subset of $\r^+$, which may be open, half-open or closed with one of the following forms:
$$[t_1, t_2], [t_1, t_2), (t_1, t_2], (t_1, t_2),$$
where $t_1,t_2\in\r^+$ and $t_2 \geq t_1$ ($t_1=t_2$ is only allowed in the case of $[t_1, t_2]$). Here, $t_1$ and $t_2$ are called the \emph{left} and \emph{right endpoints} of $\t$, respectively. Conveniently, we use $\inf\t$ and $\sup\t$ to denote $t_1$ and $t_2$, respectively. In this paper, we only consider {\bf bounded intervals}.

For reasoning about the temporal properties, we further define the \emph{addition} and \emph{subtraction} of (time) intervals. The expression $\t+t$ or $t+\t$, for $t\in\r^{+}$, denotes the interval $\{t + t': t'\in \t\}$. Similarly, $\t-t$ stands for the interval $\{-t + t': t'\in \t\}$ if $t\leq \inf\t$. Furthermore, for two intervals $\t_1$ and $\t_2$, $$\t_1+\t_2=\{t\in (t'+\t_2): t'\in \t_1\}=\{t_1+t_2: t_1\in\t_1 \textrm{ and } t_2\in \t_2\}.$$ Two intervals $\t_1$ and $\t_2$ are \emph{disjoint} if their intersection is an empty set, i.e., $\t_1\cap\t_2=\emptyset$. Let us see some concrete examples: $1+(2,3)=(3,4)$, $(2,3)-1=(1,2)$, $(2,3)+[3,4]=(5,7)$ and $(2,3),[3,4]$ are disjoint. It is obvious that all calculations of time intervals in the above are easy to be computed.

An \emph{algebraic number} is a complex number that is a root of a non-zero polynomial in one variable with rational coefficients (or equivalent to integer coefficients, by eliminating denominators). An algebraic number $\alpha$ is represented by $(P, (a, b), \epsilon)$ where $P$ is the minimal polynomial of $\alpha$, $a,b\in\q$ and $a +bi$ is an approximation of $\alpha$ such that $|\alpha - (a + bi)| < \epsilon$ and $\alpha$ is the only root of $P$ in the open ball $B(a + bi, \epsilon)$. The minimal polynomial of $\alpha$ is the polynomial with the smallest degree in $\q[t]$ such that $\alpha$ is a root of the polynomial and the coefficient of the highest-degree term is 1. Any root of $f(t)\in\a[t]$ is algebraic.
Moreover, given the representations of $a, b\in \a$, the representations of $a\pm b, \frac{a}{b}$ and $a\cdot b$ can be computed in polynomial time, so does the
equality checking~\cite{cohen2013course}.

Furthermore, a complex number is called \emph{transcendental} if it is not an algebraic number.
In general, it is challenging to verify relationships between transcendental numbers~\cite{richardson1997recognize}. On the other hand, one can use the Lindemann-Weierstrass theorem to compare some transcendental numbers. The transcendence of $e$ and $\pi$ are direct corollaries of this theorem.
\begin{theorem}[Lindemann-Weierstrass theorem]\label{Thm:LW}
Let $\eta_1, \cdots, \eta_n $ be pairwise distinct algebraic complex numbers. Then $\sum_k \lambda_k e^{\eta_k }\neq 0$ for non-zero algebraic numbers $\lambda_1,\cdots,\lambda_n$.
\end{theorem}

The following concepts are introduced to study the general relation between transcendental numbers.
\begin{definition}[Algebraic independence]
A set of complex numbers $S = \{a_1 , \cdots, a_n \}$ is \emph{algebraically independent} over $\q$ if the elements of $S$ do not satisfy any nontrivial (non-constant) polynomial equation with coefficients in $\q$.
\end{definition}
By the above definition, for any transcendental number $u$, $\{u\}$ is algebraically independent over $\q$, while $\{a\}$ for any algebraic number $a\in\a$ is not. Thus, a set of complex numbers that is algebraically independent over $\q$ must consist of transcendental numbers. $\{\pi,e^{\pi\sqrt{n}}\}$ is also algebraically independent over $\q$ for any positive integer $n$~\cite{nesterenko1996modular}. Checking the algebraic independence is challenging. For example, it is still widely open whether $\{e,\pi\}$ is algebraically independent over $\q$.

\begin{definition}[Extension field]
Given two fields $E\subseteq F$, $F$ is an extension field of $E$, denoted by $F/E$, if the operations of $E$ are those of $F$ restricted to $E$.
\end{definition}
For example, under the usual notions of addition and multiplication, the field of complex numbers is an extension field of real numbers.
\begin{definition}[Transcendence degree]
Let $L$ be an extension field of $\q$, the transcendence degree of $L$ over $\q$ is defined as the largest cardinality of an algebraically independent subset of $L$ over $\q$.
\end{definition}
For instance, let $\q(e)/\q=\{a+ b e\mid a,b\in\q\}$ and $\q(\sqrt{2})/\q=\{a+ b \sqrt{2}\mid a,b\in\q\}$ be two extension fields of $\q$. Then the transcendence degree of them are $1$ and $0$, respectively, by noting that $e$ is a transcendental number and $\sqrt{2}$ is an algebraic number.

Now, Schanuel's conjecture is ready to be presented.
\begin{conjecture}[Schanuel's conjecture]\label{Schanuel's conjecture}
Given any complex numbers $z_1 ,\cdots, z_n$ that are linearly independent over $\q$, the extension field $\q(z_1,...,z_n,e^{z_1} ,...,e^{z_n})$ has transcendence degree of at least $n$ over $\q$.
\end{conjecture}

Stephen Schanuel proposed this conjecture during a course given by
Serge Lang at Columbia in the 1960s~\cite{lang1966introduction}. Schanuel's conjecture concerns the transcendence degree of certain field extensions of the rational numbers. The conjecture, if proven, would generalize the most well-known results in transcendental number theory significantly~\cite{macintyre1996decidability,terzo2008some}. For example, the algebraical independence of $\{e, \pi\}$ would simply follow by setting $z_1 = 1$ and $z_2 = \pi i$, and using \emph{Euler's identity} $e^{\pi i}+1=0$.

\section{Continuous-time Markov Chains as Distributions
Transformers}\label{Sec:Symbolic}
We begin with the definition of \emph{continuous-time Markov chains (CTMCs)}.
A CTMC is a Markovian (memoryless) \emph{stochastic process} that takes values on a finite state set $S$ ($|S|=d<\infty$) and evolves in continuous-time $t\in\r^+$. Formally,
\begin{definition}
A CTMC is a pair $\m=(S,Q)$, where $S$ ($|S|=d$) is a finite state set and $Q\in \q^{d\times d}$ is a \emph{transition rate matrix}.
\end{definition}
A transition rate matrix $Q$ is a matrix whose off-diagonal entries $\{Q_{i,j}\}_{i\not=j}$ are nonnegative rational numbers, representing the transition rate from state $i$ to state $j$, while the diagonal entries $\{Q_{j,j}\}$ are constrained to be $-\sum_{j\not= i}Q_{i,j}$ for all $1\leq j\leq d$. Consequently, the column summations of $Q$ are all zero.

The evolution of a CTMC can be regarded as a \emph{distribution transformer}. Given initial distribution $\mu\in\q^{d\times 1}\in\ds$, the distribution at time $t\in \r^+$ is:
\[\mu_t=e^{Qt}\mu,\]
where $\ds$ is denoted as the set of all probability distributions over $S$. We call $\ds$ the \emph{probability distribution space} of CTMCs.
An \emph{execution path} of CTMCs is a continuous function indexed by initial distribution $\mu\in\ds$:
\begin{equation}\label{Eq:path}
\sigma_{\mu} \colon \mathbb{R}^+ \rightarrow \ds,\qquad \sigma_{\mu}(t)=e^{Qt}\mu.
\end{equation}

\begin{example}\label{Exa:CTMC}
We recall the illustrating example of CTMC $\m=(S,Q)$ in \cite[Figure 1]{aziz2000model} as the running example in our work. In particular, $\m$ is a 5-dimensional CTMC with initial distribution $\mu$, where $S=\{s_0,s_1,s_2,s_3,s_4\}$ and
\[Q=\left(\begin{matrix}
-3 & 0 & 0 & 0 & 0\\
1 & 0 & 0 & 0 & 0\\
2 & 0 & -7 & 0 & 0\\
0 & 0 & 3 & 0 & 0\\
0 & 0 & 4 & 0 & 0
\end{matrix}\right ) \qquad \mu=\left(\begin{matrix}
0.1 \\
0.2 \\
0.3 \\
0.4 \\
0
\end{matrix}\right ).\]
\end{example}

\section{Symbolic Dynamics of CTMCs}\label{sec:symbolic_dynamic}
In this section, we introduce symbolic dynamics to characterize the properties of the probability distribution space of CTMCs.

First, we fix a finite set of intervals $\ii=\{\i_{k}\subseteq [0,1]\}_{k\in K}$, where the endpoints of each $\i_k$ are rational numbers. With the states $S=\{s_0,s_1,\cdots,s_{d-1}\}$, we define the \emph{symbolization} of distributions as a function:
\begin{equation}\label{Eq:symbolize}
\s:\ds\rightarrow 2^{ S\times \ii} \qquad \s(\mu)=\{\pair{s}{\i}\in S\times \ii :\mu(s)\in\i\},
\end{equation}
where $\times$ denotes the Cartesian product, and $2^{ S\times \ii}$ is the power set of $S\times \ii$. $\pair{s}{\i}\in \s(\mu)$ asserts that the probability of \emph{state s} in distribution $\mu$ is in the interval $\i$. The \emph{symbolization of distributions} is a generalization of the discretization of distributions with $\i_{k}\cap\i_m=\emptyset$ for all $k\not =m$ which was studied in \cite{agrawal2015approximate}. This generalization increases the expressiveness of our \emph{continuous linear-time logic} introduced in the next section. Now, we can represent any given probability distribution by finite symbols from $S\times \ii$.
For example, suppose
\begin{equation}\label{Eq:symbolization}
\ii=\{[0, 0.1], (0.1, 0.9), [0.9, 1], [1, 1], [0.4,0.4] \},
\end{equation} and then the initial distribution $\mu$ in Example~\ref{Exa:CTMC} is symbolized as
\begin{equation}
\begin{aligned}
\s(\mu)=\{&\pair{s_0}{[0, 0.1]},\pair{s_1}{(0.1, 0.9)},\pair{s_2}{(0.1, 0.9)},\\
&\pair{s_3}{(0.1, 0.9)},\pair{s_3}{[0.4, 0.4]},\pair{s_4}{[0, 0.1]}\}.
\end{aligned}
\end{equation}
As we can see from the above example, the symbolization of distributions on states considers the exact probabilities (singleton intervals) of the states and the range of their possibilities.

Next, we introduce the symbolization to CTMCs,
\begin{definition}
A \emph{symbolized} CTMC is a tuple $\sm=(S,Q,\ii)$, where $\m=(S,Q)$ is a CTMC and $\ii$ is a finite set of intervals in $[0,1]$.
\end{definition}
As we can see, the set of intervals is picked depending on CTMCs.
Then, we extend this symbolization to the path $\sigma_\mu$:
\begin{equation}\label{Eq:symbolic_dynamics}
\s\circ\sigma_\mu:\r^+\rightarrow 2^{S\times \ii}.
\end{equation}

\begin{definition}\label{Eq:continuous_word}
Given a symbolized CTMC $\sm=(S,Q,\ii)$, $\s\circ\sigma_\mu$ is a \emph{symbolic execution path} of $\m=(S,Q)$.
\end{definition}
Given a symbolized CTMC $\sm=(S,Q,\ii)$, the path $\sigma_\mu$ of CTMC $\m=(S,Q)$ over real numbers $\r^{+}$ generated by probability distribution $\mu$ induces a symbolic execution path $\s\circ\sigma_\mu$ over finite symbols $S\times \ii$. Subsequently, the dynamics of CTMCs can be studied in terms of a language over $S\times \ii$. In other words, we can study the temporal properties of CTMCs in the context of symbolized CTMCs.

\section{Continuous Linear-time Logic}\label{Sec:CLL}In this section, we introduce \emph{continuous linear-time logic (CLL)}, a probabilistic linear-time temporal logic, to specify the temporal properties of a symbolized CTMC $\sm=(S,Q,\ii)$.

CLL has two types of formulas: state formulas and path formulas. The state formulas are constructed using propositional connectives. The path formulas are obtained by propositional connectives and a temporal modal operator \emph{timed until} $U^{\t}$ for a bounded time interval $\t$, as in MTL and CSL. Furthermore, \emph{multiphase timed until formulas} $\Phi_0U^{\t_1}\Phi_1U^{\t_2}\Phi_2\ldots U^{\t_n}\Phi_n$ are allowed to enrich the expressiveness of CLL. More importantly, \emph{time reset} is involved in these multiphase formulas. Thus absolutely and relatively temporal properties of CTMCs can be studied.

\begin{definition}
The \emph{state formulas} of CLL are described according to the following syntax:
\[\Phi:= \mathbf{true} \mid a\in \ap\mid \neg \Phi \mid \Phi_{1} \land \Phi_{2}\] where $\ap$ denotes $S\times \ii$ as the set of atomic propositions.

The \emph{path formulas} of CLL are constructed by the following syntax:

$$\vp:=\mathbf{true} \mid \Phi_0U^{\t_1}\Phi_1U^{\t_2}\Phi_2\ldots U^{\t_n}\Phi_n \mid \neg\vp\mid \vp_1\land\vp_2$$
where $n\in\z^{+}$ is a positive integer, for all $0\leq k\leq n$, $\Phi_{k}$ is a state formula, and $\t_k$'s are \emph time intervals with the endpoints in $\q^{+}$, i.e., each $\t_k$ is one of the following forms:
$$(a,b),[a,b],(a,b], [a,b) \qquad \forall a,b\in \q^+.$$
\end{definition}

The semantics of CLL state formulas is defined on the set $\ds$ of probability distributions over $S$ with the symbolized function $\s$ in Eq.(\ref{Eq:symbolize}) of Section \ref{sec:symbolic_dynamic}.

\begin{itemize}
\item [(1)] $\mu\models \mathbf{true}$ for all probability distributions $\mu\in\ds$;
\item [(2)] $\mu \models a$ iff $a\in\s(\mu)$;
\item [(3)] $\mu \models \neg \Phi$ iff it is not the case that $\mu\models \Phi$ (or written $\mu \not \models \Phi$ );
\item [(4)] $\mu \models \Phi_{1} \land \Phi_{2}$ iff $\mu \models \Phi_{1}$ and $\mu \models \Phi_{2}$.
\end{itemize}
The semantics of CLL path formulas is defined on execution paths $\{\sigma_{\mu}\}_{\mu\in\ds}$ of CTMC $\m=(S,Q)$.

\begin{itemize}
\item [(1)] $\sigma_\mu\models \mathbf{true}$ for all probability distributions $\mu\in \ds$;

\item [(2)] $\sigma_\mu \models\Phi_0U^{\t_1}\Phi_1U^{\t_2}\Phi_2\ldots U^{\t_n}\Phi_n$ iff there is a time instant $t\in\t_1$ such that $\sigma_{\mu_{t}} \models \Phi_1U^{\t_2}\Phi_2\ldots U^{\t_n}\Phi_n$, and for any $t'\in \t_1\cap[0,t)$, ${\mu_{t'}}\models \Phi_0$, where $\sigma_{\mu_t}\models \Phi$ iff $\mu_t\models \Phi$, and $\mu_{t}$ is the distribution of the chain at time instant $t$, i.e., $\mu_{t}=e^{Qt}\mu \ \forall t\in\r^{+}$;

\item [(3)] $\sigma_\mu \models \neg \vp$ iff it is not the case that $\sigma_\mu\models \vp$ (written $\sigma_\mu \not \models \vp$ );
\item [(4)]
$\sigma_\mu \models \vp_{1} \land \vp_{2}$ iff $\sigma_\mu \models \vp_{1}$ and $\sigma_\mu \models \vp_{2}$.
\end{itemize}
Not surprisingly, other Boolean connectives are derived in the standard way, i.e., $\mathbf{false}=\neg\mathbf{true}$, $\Phi_1\lor \Phi_2=\neg(\neg\Phi_1\land \neg\Phi_2)$ and $\Phi_1\rightarrow\Phi_2=\neg\Phi_1\lor\Phi_2,$ and the path formula $\vp$ follows the same way. Furthermore,
we generalize temporal operators $\Diamond$ (“eventually”) and $\Box$ (“always”) of discrete-time systems into their timed variant $\Diamond^{\t}$ and $\Box^{\t}$, respectively, in the following:
\[\Diamond^\t\Phi= \mathbf{true} U^\t\Phi \qquad\Box^\t\Phi=\neg \Diamond^\t\neg \Phi.\]

For $n=1$ in multiphase timed until formulas, the until operator $ U^{\t_1}$ is a timed variant of the until operator of LTL; the path formula $\Phi_0 U^{\t_1} \Phi_1$ asserts that $\Phi_1$ is satisfied at some time instant in the interval $\t_1$ and that at all preceding time instants in $\t_1$, $\Phi_0$ holds. For example,
$$\vp=\pair{s_1}{[0,0.1]}U^{[0,5]}\pair{s_0}{[0.9,1]},$$
as mentioned in introduction section.

For general $n$, the CLL path formula $\Phi_0U^{\t_1}\Phi_1U^{\t_2}\Phi_2\ldots U^{\t_n}\Phi_n$ is explained over the induction on $n$. We first mention that $U^\t$ is right-associative, e.g., $\Phi_0 U^{\t_1}\Phi_1 U^{\t_2}\Phi_2$ stands for $\Phi_0 U^{\t_1}(\Phi_1 U^{\t_2}\Phi_2)$. This makes time reset, i.e., $\t_1$ and $\t_2$ do not have to be disjoint, and the starting time point of $\t_2$ is based on some time instant in $\t_1$. Recall the multiphase timed until formula in introduction section and this formula expresses a relative time property:
\[\vp'=\pair{s_0}{[0.9,1]}U^{[3,7]}(\pair{s_1}{[0,0.1]}U^{[0,5]}\pair{s_0}{[0.9,1]}),\]which is different to the following CLL path formula representing an absolutely temporal property of CTMCs:
\[\vp''=\Box^{[3,7]}\pair{s_0}{[0.9,1]}\land \pair{s_1}{[0,0.1]}U^{[0,5]}\pair{s_0}{[0.9,1]}).\]
As an example, we clarify the semantics of CLL by comparing the above two path formulas in general forms:
\[\Phi_0U^{\t_1}\Phi_1 U^{\t_2}\Phi_2 \ \text{ and }\ \Phi_0U^{\t_1}\Phi_1\land \Phi_1 U^{\t_2}\Phi_2.\]

\begin{itemize}
\item [(1)] $\sigma_\mu\models \Phi_0U^{\t_1}\Phi_1 U^{\t_2}\Phi_2$ asserts that there are time instants $t_1\in\t_1, t_2\in \t_2$ such that $\mu_{t_1+t_2}\models \Phi_2$ and for any $t_1'\in\t_1\cap [0,t_1)$ and $t_2'\in \t_2\cap[0,t_2)$, $\mu_{t_1'}\models \Phi_0$ and $\mu_{t_1+t_2'}\models \Phi_1$, where $\mu_t=e^{Qt}\mu \ \forall t\in\r^{+}.$
This is more clear in the following timeline.

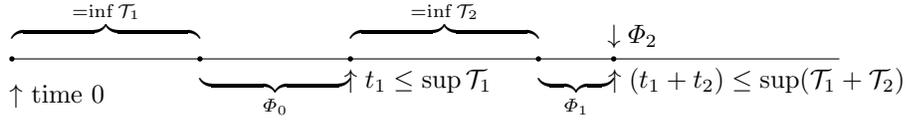
\begin{tikzpicture}
\draw (-4.5,0)--(0,0)--(1,0)--(2,0)--(3,0)--(4,0)--(6,0);
\fill (-0.5,0) circle (1pt);
\fill (2,0) circle (1pt);
\fill (3,0) circle (1pt);
\fill (-2.5,0) circle (1pt);
\fill (-4.5,0) circle (1pt);
\node[rotate = 0] at (-1.5, -0.5) {$\underbrace{\hspace{2cm}}_{\Phi_{0}}$};
\node[rotate = 0] at (2.5, -0.5) {$\underbrace{\hspace{1cm}}_{\Phi_{1}}$};
\node[rotate = 0] at (0.75, 0.5) {$\overbrace{\hspace{2.5cm}}^{= \inf\t_2}$};
\node[rotate = 0] at (3.25, 0.3) {$\downarrow{\Phi_{2}}$};
\node[rotate = 0] at (-3.5, 0.5) {$\overbrace{\hspace{1.9cm}}^{= \inf\t_1}$};
\node[rotate = 0] at (-3.95, -0.3) {$\uparrow$ time 0};
\node[rotate = 0] at (0.4, -0.3) {$\uparrow t_1\leq \sup\t_1$ };
\node[rotate = 0] at (4.9, -0.3) {$\uparrow (t_1+t_2)\leq \sup(\t_1+\t_2)$ };
\end{tikzpicture}

\item [(2)] $\sigma_\mu\models \Phi_0U^{\t_1}\Phi_1\land \Phi_1 U^{\t_2}\Phi_2$ asserts that there are time instants $t_1\in\t_1, t_2\in \t_2$ such that $\mu_{t_1}\models \Phi_1$ and $\mu_{t_2}\models \Phi_2$, and for any $t_1'\in\t_1\cap [0,t_1)$ and $t_2'\in \t_2\cap[0,t_2)$, $\mu_{t_1'}\models \Phi_0$ and $\mu_{t_2'}\models \Phi_1$, where $\mu_t=e^{Qt}\mu \ \forall t\in\r^{+}$.
\end{itemize}

Before solving the model-checking problem of CTMCs against CLL formulas in the next section, we shall first discuss what can be specified in our logic CLL.

Given a CTMC $(S,Q)$, CLL path formula $\Diamond^{[0,1000]}\pair{s}{[1,1]}$ expresses a liveness property that \emph{state }$s \in S$ is eventually reached with probability one before time instant $1000$. In terms of safety properties, formula $\Box^{[100,1000]}\pair{s}{[0,0]}$ represents that state $s\in S$ is never reached (reached with probability zero) between time instants $100$ and $1000$. Furthermore, setting the intervals nontrivial (neither $[0,0]$ or $[1,1]$), liveness and safety properties can be asserted with probabilities, such as $\Diamond^{[0,1000]}\pair{s}{[0.5,1]}$ and $\Box^{[100,1000]}\pair{s}{[0,0.5]}$. For multiphase timed until formula $\pair{s}{[0.7,1]}U^{[2,3]}\pair{s}{[0.7,1]}\ldots U^{[2,3]}\pair{s}{[0.7,1]},$
where the number of $U^{[2,3]}$ is 100, asserts that the probability of \emph{state} $s$ is beyond $0.7$ in every time instant $2$ to $3$, and this happens at least 100 times.

Next, we can classify members of $\ii$ as representing ``low'' and ``high'' probabilities. For example, if $\ii$ contains 3 intervals $\{[0,0.1],(0.1,0.9),[0.9,1]\}$, we can declare the first interval as ``low'' and the last interval as ``high''. In this case $\Box^{[10,1000)}(\pair{s_0}{[0,0.1]}\rightarrow \pair{s_1}{[0.9,1]})$ says that, in time interval $[10,1000)$, whenever the probability of state $s_0$ is low, the probability of state $s_1$ will be high.

\section{CLL Model Checking}\label{Sec:CLLmodelchecking}
In this section, we provide an algorithm to model check CTMCs against CLL formulas, i.e., the following CLL model-checking problem --- Problem~\ref{Problem:model_checking} is decidable.
\begin{problem}[CLL Model-checking Problem]\label{Problem:model_checking}
Given a symbolized CTMC $\sm = (S, Q, \ii)$ with an initial distribution $\mu$ and a CLL path formula $\vp$ on $\ap=S\times \ii$, the goal is to decide whether $\sigma_\mu \models \vp$, where $\sigma_\mu(t)=e^{Qt}\mu$ is an execution path defined in~Eq.(\ref{Eq:path}). 
\end{problem}
In particular, we show that
\begin{theorem}\label{theo:main}
Under the condition that Schanuel's conjecture holds, the CLL model-checking problem in Problem~\ref{Problem:model_checking} is decidable.
\end{theorem}

In the following, we prove
the above theorem from
checking basic formulas --- atomic propositions to the most complex one --- nontrivial multiphase timed until formulas. For readability, we put the proofs of all results in Appendix~\ref{Appendix:proof}.

We start with the simplest case of atomic proposition $\pair{s}{\i}$. By the semantics of CLL, $\mu_t\models \pair{s}{\i}$ if and only if $\mu_t=e^{Qt}\mu(s)\in \i$. To check this, we first observe that the execution path $e^{Qt}\mu$ of CTMCs is a system of \emph{polynomial exponential functions (PEFs)}.
\begin{definition}
A function $f:\reals\to\reals$ is a \emph{polynomial-exponential function (PEF) }if $f$ has the following form:
\begin{equation}\label{Eq:PEF}
f(t)=\sum_{k=0}^{K}f_k(t)e^{\lambda_k t}
\end{equation}
where for all $0\leq k\leq K<\infty$, $ f_k(t)\in\f_1[t],f_k(t)\not =0,$ $\lambda_k\in\f_2$ and $\f_1, \f_2$ are fields.
Without loss of generality, we assume that  $\lambda_k$'s are distinct.
\end{definition}

Generally, for a PEF $f(t)$ with the range in complex numbers $\c$, $g(t)=f(t)+f^*(t)$ is a PEF with the range in real numbers $\r$, where $f^{*}(t)$ is the complex conjugate of $f(t)$. The factor $t$ is omitted whenever convenient, i.e., $f=f(t)$. $t$ is called a \emph{root} of a function $f$ if $f(t)=0$. PEFs often appear in transcendental number theory as auxiliary functions in the proofs involving the exponential function~\cite{baker1990transcendental}.

\begin{lemma}\label{Lem:PEF_express}
Given a CTMC $\m=(S,Q)$ with $S=\{s_0,\ldots,s_{d-1}\}$, $Q\in \mathbb{Q}^{d\times d}$, and an initial distribution $\mu\in \q^{d\times 1}$, for any $0\leq i\leq d-1$ , $e^{Qt}\mu (s_i)$, the $i$-th entry of $e^{Qt}\mu$, can be expressed as a PEF $f:\r^+\rightarrow [0,1]$ as in Eq.(\ref{Eq:PEF}) with $\f_1=\f_2=\a$.
\end{lemma}
By the above lemma, for a given $t$ in some bounded time interval $\t$ (to be specific in the latter discussion), $e^{Qt}\mu(s)\in \i$ is determined by the algebraic structure of PEF $g(t)=e^{Qt}\mu(s)$ in $\t$. That is all \emph{maximum intervals} $\t_{\max}\subseteq\t$ such that $g(t)\in \i$ for all $t\in\t_{\max}$, where interval $\t_{\max}\not =\emptyset$ is called maximum for $g(t)\in \i$ if no sub-intervals $\t'\subsetneq\t_{\max} $ such that the property holds, i.e., $g(t)\in \i$ for all $t\in\t'$. Then $e^{Qt}\mu(s)\in \i$ if and only if $t\in\t_{\max}$ for some maximum interval $\t_{\max}$. So, we aim to compute the set $\tt$ of all maximum intervals. By the continuity of PEF $g(t)$, this can be done by identifying a \emph{real root isolation} of the following PEF $f(t)$ in $\t$: $f(t)=(g(t)-\inf\i)(g(t)-\sup\i).$

A (real) \emph{root isolation} of function $f(t)$ in interval $\t$ is a set of mutually disjoint intervals, denoted by $\iso(f)_\t=\{ (a_j,b_j)\subseteq \t\}$ for $a_j,b_j\in\q$ such that \begin{itemize}
\item for any $j$, there is one and only one root of $f(t)$ in $(a_j,b_j)$;
\item for any root $t^*$ of $f(t)$, $t^*\in (a_j,b_j)$ for some $j$.
\end{itemize}
Furthermore, if $f$ has no any root in $\t$, then $\iso(f)_\t=\emptyset$.

Although there are infinite kinds of real root isolations of $f(t)$ in $
\t$, the number of isolation intervals equals to the number of distinct roots of $f(t)$ in $\t$.

Finding real root isolations of PEFs is a long-standing problem and can be at least backtracked to Ritt's paper~\cite{ritt1929zeros} in 1929. Some following results were obtained since the last century (e.g.~\cite{avellar1980zeros,tijdeman1971number}). This problem is essential in the reachability analysis of dynamical systems, one active field of symbolic and algebraic computation. In the case of $\f_1=\q$ and $\f_2=\mathbb{N}^+$ in \cite{achatz2008deciding}, an algorithm named ISOL was proposed to isolate all real roots of $f(t)$. Later, this algorithm has been extended to the case of $\f_1=\q$ and $\f_2=\r$~\cite{gan2017reachability}. A variant of the problem has also been studied in \cite{li2016positive}. The correctness of these algorithms is based on Schanuel's conjecture. Other works are using Schanuel's conjecture to do the root isolation of other functions, such as exp-log functions \cite{strzebonski2008real} and tame elementary functions \cite{strzebonski2009real}.

By Lemma~\ref{Lem:PEF_express}, we pursue this problem in the context of CTMCs. The distinct feature of solving real root isolations of PEFs in our paper is to deal with complex numbers $\c$, more specifically algebraic numbers $\a$, i.e., $\f_1=\f_2=\a$. At the same time, to the best of our knowledge, all the previous works can only handle the case over $\r$. Here, we develop a state-of-the-art real root isolation algorithm for PEFs over algebraic numbers. Thus
from now on, we always assume that PEFs are over $\a$, i.e., $\f_1=\f_2=\a$ in Eq.(\ref{Eq:PEF}). In this case, it is worth noting that whether a PEF has a root in a given interval, $\t\subseteq \r^{+}$ is decidable subject to Schanuel's Conjecture if $\t$ is bounded~\cite{chonev2015skolem}, which falls in the situation we consider in this paper.
\begin{theorem}[\cite{chonev2015skolem}]\label{Thm:existence}
Under the condition that Schanuel's conjecture holds, there is an algorithm to check whether a PEF $f(t)$ has a root in interval $\t$, i.e., whether $\iso(f)_\t=\emptyset$.
\end{theorem}

In this paper, we extend the above checking $\iso(f)_\t=\emptyset$ to computing $\iso(f)_\t$ of PEF $f(t)$.
\begin{theorem}\label{Thm:isolating}
Under the condition that Schanuel's conjecture holds, there is an algorithm to find real root isolation $\iso(f)_\t$ for any PEF $f(t)$ and interval $
\t$. Furthermore, the number of real roots is finite, i.e., $|\iso(f)_\t|<\infty$.
\end{theorem}

We can compute the set $\tt$ of all maximum intervals with the above theorem to check atomic propositions. Furthermore,
we can compare the values of any real roots of PEFs, which is important in model checking general multiphase timed until formulas at the end of this section.
\begin{lemma}\label{lem:compare}
Let $f_1(t)$ and $f_2(t)$ be two PEFs with the domains in $\t_1$ and $\t_2$, and $t_1\in \t_1$ and $t_2\in \t_2$ are roots of them, respectively. Under the condition that Schanuel's conjecture holds, there is an efficient way to check whether or not $t_1-t_2<g$ for any given rational number $g\in\q$.
\end{lemma}

For model checking general state formula $\Phi$, we can also use real root isolation of some PEF to obtain the set of all maximum intervals $\t_{\max}$ such that $\mu_t\models \Phi$ for all $t\in\t_{\max}$. The reason is that $\Phi$ admits \emph{conjunctive normal form} consisting of atomic propositions. See the proof of the following lemma in Appendix~\ref{Appendix:proof}.
\begin{lemma}\label{Lem:ap_compute}
Under the condition that Schanuel's conjecture holds, given a time interval $\t$, the set $\tt$ of all maximum intervals in $\t$ satisfying $\mu_t\models\Phi$ can be computed, where $\Phi$ is a state formula of CLL. Furthermore,
the number of all intervals in $\tt$ is finite; the left and right endpoints of each interval in $\tt$ are roots of PEFs.
\end{lemma}

At last, we characterize the multiphase timed until formulas by the reachability analysis of time intervals (instants).

\begin{lemma}\label{Lem:characterization}
$\sigma_\mu \models \Phi_0U^{\t_1}\Phi_1U^{\t_2}\Phi_2\cdots U^{\t_n}\Phi_{n}$ if and only if there exist time intervals $\{\i_{k}\subseteq \r^{+}\}_{k=0}^n$ with $\i_0=[0,0]$ such that
\begin{itemize}
\item \emph{The satisfaction of intervals}: for all $1\leq k\leq n$, ${\mu_t}\models \Phi_{k-1}$ for all $t\in\i_k$, and ${\mu_{t^*}}\models \Phi_{n}$, where $t^*=\sup\i_n$ and $\mu_t=e^{Qt}\mu\ \forall t\in\r^{+}$;
\item \emph{The order of intervals}: for all $1\leq k\leq n$, $\i_k\subseteq \i_{k-1}+\t_k$ and $\inf\i_k=\sup\i_{k-1}+\inf\t_k$.
\end{itemize}
\end{lemma}

By the above lemma, the problem of checking multiphase timed until formulas is reduced to verify the existence of a sequence of time  intervals.

Now we can show the proof of Theorem~\ref{theo:main}.

\begin{proof}
Recall that the nontrivial step is to model check multiphase timed until formula $\Phi_0U^{\t_1}\Phi_1U^{\t_2}\Phi_2\cdots U^{\t_n}\Phi_{n},$ where $\{\t_{j}\}_{j=1}^n$ is a set of bounded rational intervals in $\r^+$, and for $0\leq k\leq n+1$, $\Phi_k$ is a state formula.

By Lemma~\ref{Lem:characterization}, for model checking the above formula, we only need to check the existence of time  intervals $\{\i_{k}\}_{k=0}^n$ illustrated in the lemma. The following procedure can construct such a set of intervals if it exists:
\begin{itemize}
\item (1) Let $\ii_0=\{\i_0=[0,0]\}$ ;
\item (2) For each $1\leq k\leq n$, obtaining the set $\ii_k$ in $[0,\sum_{j=1}^{k}\sup\t_j]$ of all maximum intervals such that $\mu_t\models \Phi_{k-1}$ for all $t\in \i$ of $\i\in\ii$, where $\mu_t=e^{Qt}\mu$; this can be done by Lemma~\ref{Lem:ap_compute}. Noting that $\ii_k$ can be the empty set, i.e., $\ii_k=\emptyset$;
\item (3) Let $k$ from $1$ to $n$. First, updating $\ii_k$:\begin{equation}
\begin{aligned}
\ii_{k}=\{\i
\cap (\i'+\t_{k}): \i\in \ii_{k}\text{ and }\i'\in\ii_{k-1}\}.
\end{aligned}
\end{equation}

The above updates can be finished by Lemma~\ref{lem:compare}. If $\ii_k=\emptyset$, then the formula is not satisfied;
\item (4) Updating $\ii_{n}$: for each $\i\in \ii_n$, we replace $\i$ with $[s-\epsilon,s)$ for some constant $\epsilon>0$ if there is an $s\in \i$ with $s-\epsilon\in\i$ such that $\mu_{s}\models \Phi_{n}$
where $\mu_{s}=e^{Qs}\mu$; Otherwise, remove this element from $\ii_{n}$. Again, this can be done by Lemma~\ref{Lem:ap_compute}. If $\ii_n=\emptyset$, then the formula is not satisfied;
\item (5) Finally, let $k$ from $n-1$ to $1$, updating $\ii_{k}$:
\[\ii_{k}=\{[s-\inf\t_{k},s-\inf\t_k]:[s-\epsilon,s)\in\ii_{k+1}]\}.\]
\end{itemize}
Thus after the above procedure, we have non-empty sets $\{\ii_{k}\}_{k=0}^{n}$ with the following properties.
\begin{itemize}
\item for each $1\leq k\leq n$, $\mu_t\models \Phi_{k-1}$ for all $t\in \i_k$ and $\i_k\in\ii_k$, and $\mu_{t^*}\models \Phi_{n}$, where $t^*=\sup\i_n$;
\item for each $1\leq k\leq n$, $\i\in \ii_k$, there exists at least one $\i'\in\ii_{k-1}$ such that $\i\subseteq \sup\i'+\t_k$ and $\inf\i=\sup\i'+\inf\t_k$.
\end{itemize}

Therefore, we can get a set of intervals $\{\i_{k}\}_{k=0}^n$ satisfying the two conditions in Lemma~\ref{Lem:characterization} if it exists. On the other hand, it is easy to check that all such $\{\i_{k}\}_{k=0}^n$ must be in $\{\ii_{k}\}_{k=0}^{n}$, i.e., for each $k$, $\i_k\subseteq \i$ for some $\i\in \ii_k$. This ensures the correctness of the above procedure.
\end{proof}

By the above constructive analysis, we give an algorithm for model checking CTMCs against CLL formulas. Focusing on the decidability problem, we do not provide the pseudocode of the algorithm. Alternatively, we implement a numerical experiment to illustrate the checking procedure in the next section.

\section{Numerical Implementation}\label{sec:case-study}
In this section,  we implement a case study of checking CTMCs against CLL formulas. Here, we consider a symbolized CTMC $\sm=(S,Q,\ii)$, where $\m=(S,Q)$ is the CTMC in Example~\ref{Exa:CTMC} and finite set $\ii$ is the one considered in Eq.(\ref{Eq:symbolization}). 
We check the properties of $\m$ given by the following two CLL path formulas mentioned in the introduction for different initial distributions. 
\begin{equation*}
    \begin{aligned}
    \vp&=\pair{s_1}{[0,0.1]}U^{[0,5]}\pair{s_0}{[0.9,1]}.\\
    \vp'&=\pair{s_0}{[0.9,1]}U^{[3,7]}\pair{s_1}{[0,0.1]}U^{[0,5]}\pair{s_0}{[0.9,1]}.
    \end{aligned}
\end{equation*}

By Jordan decomposition, we have 
$Q=SJS^{-1}$ where 
 \[S=\left(\begin{matrix}
  0 & -6 & 0 & 0 & 0\\
  0  & 2 & 0 & 0 & 1\\
  -7  & -3 & 0 & 0 & 0\\
  3  & 3 & 0 & 1 & 0\\
  4  & 4 & 1 & 0 & 0
  \end{matrix}\right ) \qquad J=\left(\begin{matrix}
  -7 & 0 & 0 & 0 & 0\\
  0  & -3 & 0 & 0 & 0\\
  0  & 0 & 0 & 0 & 0\\
  0  & 0 & 0 & 0 & 0\\
  0  & 0 & 0 & 0 & 0
  \end{matrix}\right ) \qquad S^{-1}=\left(\begin{matrix}
  \frac{1}{14} & 0 & -\frac{1}{7} & 0 & 0\\
  -\frac{1}{6}  & 0 & 0 & 0 & 0\\
  \frac{8}{21}  & 0 & \frac{4}{7} & 0 & 1\\
  \frac{2}{7}  &0  & \frac{3}{7} & 1 & 0\\
  \frac{1}{3}  &1  & 0 & 0 & 0
  \end{matrix}\right ).\] 
  Then, we consider an initial distribution $\mu$ as the same as the one in Example~\ref{Exa:CTMC}. Then we have that the value of $e^{Qt}\mu$ is as follows:
 \[
  \left(\begin{matrix}
  e^{-3t}&  0 \qquad &  0 \qquad & 0\qquad & 0  \\

  -\frac{1}{3}(e^{-3t}-1) & 1  & 0 & 0 & 0\\

  \frac{1}{2}(e^{-3t}-e^{-7t})  & 0 & e^{-7t} & 0 & 0\\

  \frac{3}{14}e^{-7t}-\frac{1}{2}e^{-3t}+\frac{2}{7} & 0 & -\frac{3}{7}e^{-7t}+\frac{3}{7} & 1 & 0\\

  \frac{2}{7}e^{-7t}-\frac{2}{3}e^{-3t}+\frac{8}{21} & 0 & -\frac{4}{7}e^{-7t}+\frac{4}{7} &  0 & 1
  \end{matrix}\right )\left(\begin{matrix}
  0.1 \\
  0.2 \\
  0.3 \\
  0.4 \\
  0
  \end{matrix}\right)=\left(\begin{matrix}
  \frac{1}{10}e^{-3t} \\
  -\frac{1}{30}e^{-3t}+\frac{7}{30}\\
  \frac{1}{20}e^{-3t}+\frac{1}{4}e^{-7t}\\
  -\frac{1}{20}e^{-3t}-\frac{3}{28}e^{-7t}+\frac{39}{70} \\
  -\frac{1}{15}e^{-3t}-\frac{1}{7}e^{-7t}+\frac{22}{105}
  \end{matrix}\right ).\]
  As we only consider states $s_0$ and $s_1$ in formulas $\vp$ and $\vp'$, we focus on the following PEFs: $		f_0(t)=\frac{1}{10}e^{-3t}$ and $f_1(t)= -\frac{1}{30}e^{-3t}+\frac{7}{30}$.
 
Next, we initialize the model checking procedures introduced in the proof of Theorem~\ref{theo:main}. First, we compute the set $\tt$ of all maximum intervals $\t\subseteq [0,5]$ such that $e^{Qt}\mu\models \pair{s_0}{[0.9,1]}$ for $t\in \t$, i.e., $f_0(t)\in [0.9,1]$ for $t\in \t$. We obtain $\tt=\emptyset$ by the real root isolation algorithm mentioned in Theorem~\ref{Thm:isolating}, and this indicates that 
$\sigma_{\mu}\not\models\vp$ 
where $\sigma_{\mu}(t)=e^{Qt}\mu$  is the  path induced by $\mu$ and defined in~Eq.(\ref{Eq:path}).

To check whether $\sigma_{\mu}\models\vp'$, we compute the set $\tt$ of all maximum intervals $\t\subseteq [0,12]$ such that $e^{Qt}\mu\models \pair{s_0}{[0.9,1]}$ for $t\in \t$, i.e., $f_0(t)\in [0.9,1]$ for $t\in \t$. Again, we obtain $\tt=\emptyset$ by the real root isolation algorithm in Theorem~\ref{Thm:isolating}. Therefore, $\sigma_{\mu}\not\models\vp'$.

In the following, we consider a different initial distribution $\mu_1$ as follows:
  \[
  e^{Qt}\mu_1=e^{Qt}\left(\begin{matrix}
  0.9 \\
  0 \\
  0.1 \\
  0 \\
  0
  \end{matrix}\right)=\left(\begin{matrix}
  \frac{9}{10}e^{-3t} \\
  -\frac{3}{10}(e^{-3t}-1)\\
  \frac{9}{20}e^{-3t}-\frac{7}{20}e^{-7t}\\
  -\frac{9}{20}e^{-3t}+\frac{3}{20}e^{-7t}+\frac{3}{10} \\
  -\frac{3}{5}e^{-3t}+\frac{1}{5}e^{-7t}+\frac{2}{5}
  \end{matrix}\right ).\]
 The key PEFs are: $g_0(t)=\frac{9}{10}e^{-3t}$ and $g_1(t)=-\frac{3}{10}(e^{-3t}-1).$

 Again, we initialize the model checking procedures introduced in the proof of Theorem~\ref{theo:main}. We first compute the set $\tt$ of all maximum intervals $\t\subseteq [0,5]$ such that $e^{Qt}\mu_1\models \pair{s_1}{[0,0.1]}$ for $t\in \t$, i.e., $g_1(t)\in [0,0.1]$ for $t\in \t$. This can be done by finding  a real root isolation of the following PEF: 
 $g_1^0(t)=-\frac{3}{10}(e^{-3t}-1)-\frac{1}{10}.$

  By implementing the real root isolation algorithm in Theorem~\ref{Thm:isolating}, we have 
  \[\iso{(g_1^0)}_{[0,5]}=\{(0.13,0.14)\}
\text{ and then } 
\tt=\{[0,t^*] \} \text{ for } t^*\in (0.13,0.14).\]
  Following the same way, we compute $\tt$ for  $e^{Qt}\mu_1\models \pair{s_0}{[0.9,1]}$. Then we complete the model checking procedures in the proof of Theorem~\ref{theo:main}, and  we conclude:
$\sigma_{\mu_1}\models\vp$. By repeating these, the result of the second formula $\vp'$ is $\sigma_{\mu_1}\not\models\vp'.$

\section{Related Works}\label{Sec:related_works}
Agrawal \textit{et al.} \cite{agrawal2015approximate} introduced probabilistic linear-time temporal logic (PLTL) to reason about discrete-time Markov chains in the context of distribution transformers as we did for CTMCs in this paper. Interestingly, the Skolem Problem can be reduced to the model checking problem for the logic PLTL~\cite{Akshay2015}. The Skolem Problem asks whether a given linear recurrence sequence has a zero term and plays a vital role in the reachability analysis of linear dynamical systems. Unfortunately, the decidability of the problem remains open~\cite{ouaknine2012decision}. Recently, the Continuous Skolem Problem has been proposed with good behavior (the problem is decidable) and forms a fundamental decision problem concerning reachability in continuous-time linear dynamical systems~\cite{chonev2015skolem}. Not surprisingly, the Continuous Skolem Problem can be reduced to model-checking CLL. The primary step of verifying CLL formulas is to find a real root isolation of a PEF in a given interval. Chonev, Ouaknine and Worrell reformulated the Continuous Skolem Problem in terms of whether a PEF has a root in a given interval, which is decidable subject to Schanuel's conjecture~\cite{chonev2015skolem}. An algorithm for finding root isolation can also answer the problem of checking the existence of the roots of a PEF. However, the reverse does not work in general. Therefore, the decidability of the Continuous Skolem Problem cannot be applied to establish that of our CLL model checking.

\begin{remark}
By adopting the method in this paper, we established the decidability of model checking quantum CTMCs against signal temporal logic~\cite{xu_et_al:LIPIcs.CONCUR.2021.13}. Again, we need Schanuel's conjecture to guarantee the correctness. A Lindblad’s master equation governs a quantum CTMC and a more general real-time probabilistic Markov model than a CTMC, i.e., a CTMC is an instance of quantum CTMCs. We converted the evolution of Lindblad’s master equation into a distribution transformer that preserves the laws of quantum mechanics. We reduced the model-checking problem of quantum CTMCs to the real root isolation problem, which we considered in this paper, and thus our method could be applied to it.
\end{remark}
\section{Conclusion}\label{Sec:Conclusion}
This paper revisited the study of temporal properties of finite-state CTMCs by symbolizing the probability value space $[0, 1]$ into a finite set of intervals. To specify relatively and absolutely temporal properties, we propose a probabilistic logic for CTMCs, namely continuous linear-time logic (CLL). We have considered the model checking problem in this setting. Our main result is that a state-of-the-art real root isolation algorithm over the field of algebraic numbers was proposed to establish the decidability of the model checking problem under the condition that Schanuel's conjecture holds.

This paper aims to show decidability in as simple a fashion as possible without paying much attention to complexity issues. Faster algorithms on our current constructions would significantly improve from a practical standpoint.

\section*{Acknowledgments}
We want to thank Professor Joost-Pieter Katoen for  his invaluable feedback and for pointing out the references~\cite{chen2011model,chen2011observing,majumdar2020decidability}. This work is supported by the National Key R$\&$D Program of China (Grant No: 2018YFA0306701), the National Natural Science Foundation of China (Grant No: 61832015), ARC Discovery Program (\#DP210102449) and ARC DECRA (\#DE180100156).
\bibliographystyle{unsrt} 
\bibliography{note110418}
\newpage
\clearpage
\appendix
\section{Appendices: Proofs}\label{Appendix:proof}
\subsection{Proof of Lemma~\ref{Lem:PEF_express}}
We need to use the \emph{Jordan decomposition} to prove Lemma~\ref{Lem:PEF_express}.
\begin{definition}
A \emph{Jordan block} is a square matrix of the following form.
$$
\left[\begin{matrix}
\lambda&1&0&\cdots &0\\
0&\lambda&1&\cdots &0\\
& &\ddots &\ddots &\\
& & &\ddots & 1\\

0 & 0 & 0 &\cdots &\lambda
\end{matrix}\right].$$
A square matrix $J$ is in Jordan norm form if
$$J=\left[\begin{matrix}
J_1&&& &\\
&J_2& & &\\
& & &\ddots &\\
&&& &J_n
\end{matrix}\right],$$
where $J_k$ is a Jordan block for each $1\leq k\leq n$.
\end{definition}
Because $\a$ is algebraic closed, we know that
\begin{proposition}[\cite{chen2015continuous}]\label{Prop:jordan}
Any matrix $A\in \q^{n\times n}$ is algebraically similar to a matrix in Jordan normal form over the algebraic number field $\a$. Namely, there exists some invertible $P\in \a^{n\times n}$ and $J \in \a^{n\times n}$ in Jordan form such that $A = P^{-1}J P$, where $\a^{n\times n}$ is the set of all $n$-by-$n$ matrices with every entry being algebraic numbers. %where $J_k$ has the form
%$$
%J_k=\left[\begin{matrix}
%$ \lambda_k&1&0&\cdots &0\\
%0&\lambda_k&1&\cdots &0\\
% & & &\ddots &\\
%0&0&0&\cdots &\lambda_k
%\end{matrix}\right]$$
\end{proposition}

Now we can give the proof of Lemma~\ref{Lem:PEF_express}.
\begin{proof}
As the elements of $\mu$ are rational, we only need to prove that any entry of $e^{Qt}$ can be expressed as a finite sum of $\sum_{k}f_k(t) e^{\eta_k t}$ for $\eta_k\in \a$ and $f_k(t)\in \a[t]$.

By Proposition~\ref{Prop:jordan}, we have that there is a $P\in \a^{n\times n}$ such that $Q=P^{-1} (\oplus_k J_k) P$ such that
$$
J_k=\left[\begin{matrix}
\lambda_k&1&0&\cdots &0\\
0&\lambda_k&1&\cdots &0\\
& & &\ddots &\\
0&0&0&\cdots &\lambda_k,
\end{matrix}\right]$$
where $\lambda_k$ is an eigenvalue of $Q$ and $Q\in \q^{d\times d}$, so $\lambda_k$ is algebraic. Furthermore, $J_k\in \a^{n_k\times n_k}$, where $n_k$ is the dimension of $J_k$.

Therefore, $e^{Qt}=P^{-1}e^{\oplus_k J_k t}P=P^{-1}(\oplus_k e^{ J_k t})P$. We complete the proof by proving that for each $k$, any entry of $e^{J_k t}$ can be expressed as a finite sum of $\sum_{k}f_k(t) e^{\eta_k t}$ for $\eta_k\in \a$ and $f_k(t)\in \a[t]$. Computing $e^{J_k t}$, we obtain that
$$
e^{J_k t}=\left[\begin{matrix}
e^{\lambda_k t}&te^{\lambda_k t}&t^2 e^{\lambda_k t} /{2!}&\cdots &t^n_k e^{\lambda_k t}/{n_k!}\\
0&e^{\lambda_k t}&te^{\lambda_k t}&\cdots &t^{n_k-1} e^{\lambda_k t}/{(n_k-1)!}\\
& & & \ddots&\\
0&0&0&\cdots &e^{\lambda_k t}

\end{matrix}\right].$$

%Furthermore, as $\lambda_k$ is the eigenvalue of $Q$, $R(\lambda_k)\leq 0$ and $R(\lambda_k)=0$ if and only if $\lambda_k=0$. These properties can be found in textbooks (e.g.\cite{norris_1997}).

\end{proof}
\subsection{Proof of Lemma~\ref{lem:compare}}

Before proving, we need the following fact observed in \cite{aziz2000model} according to the Lindemann-Weierstrass theorem.
\begin{observation}\label{Pro:compare}
Given a real number $r\in\r$ of the form $\sum_{k}\mu_k e^{\eta_k }$ where ${\mu_k}'s$
and ${\eta_k}'s$ are algebraic complex numbers, and the ${\eta_k}'s$ are pairwise distinct, there is an effective procedure to compare the values of $r$ and $c$ for any $c\in \q$.
\end{observation}
According to Lindemann-Weierstrass theorem, we know that $r=c$ if and only if $r=c=0$ or $c=-\mu_k$ for some $k$ with $\eta_k=0$. Otherwise, we can compute a good approximation of $r$. For each $k$, $e^{\eta_k}$ can be approximated with an error, the norm of the difference between $r$ and the approximation, of less than $\epsilon$ (for any $\epsilon<1$) by taking the first $\lceil 3|\eta_k|^2/\epsilon\rceil+1$ terms of the Maclaurin expansion for $e^{\eta_k}$. This leads to an approximation of $r$ within $\sum_{k} |\mu_k|\epsilon$. Since the individual terms in the Maclaurin expansion are algebraic functions of the $\eta_k$'s, it follows that the approximations are algebraic. Then we can check if $r>c$ by the comparision between the approximations and $c$. See~\cite{aziz2000model} for more details.

\begin{proof}
First, by Theorem~\ref{Thm:isolating}, isolating the real roots of $f_1(t)$ and $f_2(t)$, we have $t_1\in (a_1,b_1)$ and $t_2\in(a_2,b_2)$ for $a_1,a_2,b_1,b_2\in \q$. Then we first check if $t_1-t_2=g$. Note that $t_1-t_2=g$ if and only if $f_1^2(t)+f_2^2(t+g)=0$ has a root in $(a_1-g,b_1-g)\cap (a_2,b_2)$. $f_1^2(t)+f_2^2(t+g)$ is still a PEF, then we can check whether or not there is a root of it in $(a_1,b_1)\cup (a_2,b_2)$ by Theorem~\ref{Thm:existence}.

If $t_1-t_2\not=g$, we answer whether or not $t_1-(t_2+g)<0$ by narrowing $(a_1,b_1)$ and $(a_2,b_2)$ and maintaining the roots of $f_1(t)$ and $f_2(t)$ in the intervals. This can be done as
we can arbitrarily narrow the interval $(a_1,b_1)$ by comparing the signs ($>0$, $<0$ or $=0$) of $f(\frac{a_1+b_1}{2})$ and $f(a_1)$, and the same way works for narrowing $(a_2,b_2)$. The sign of $f(a)$ for any $a\in\q$ can be obtained by Observation~\ref{Pro:compare} and Theorem~\ref{Thm:LW}. Moreover, there is a gap between $t_{1}$ and $t_2+g$, and we can compare $t_1$ and $(t_2+g)$ by continuously narrowing $(a_1,b_1)$ and $(a_2,b_2)$, and comparing $a_1$ and $b_2$ ($a_2$ and $b_1$).
\end{proof}

\subsection{Proof of Lemma~\ref{Lem:ap_compute}}

\begin{proof}
Recall that state formulas of CLL are given by the following grammar:
\[\Phi:= \mathbf{true} \mid a\in \ap\mid \neg \Phi \mid \Phi_{1} \land \Phi_{2}.\]
By the semantics of CLL state formulas, $\Phi_k$ is a formula of propositional logics. So $\Phi$ admits \emph{conjunctive normal form (CNF)}~\cite[Appendix A.3]{baier2008principles}, i.e.,
$$\Phi=\land_{j\in J}\lor_{ l \in L_{j}}lit_{j,l},$$
where $lit_{j,l}$ is a \emph{literal} of $a\in\ap$ or $\neg a$, and $J_k$ and $L_k$ are both finite sets. Furthermore, we observe that $\neg a$ is (semantically equivalent to) either an atomic proposition or a disjunction of two atomic propositions. To prove this, let $a=\pair{k}{\t}$ for some interval $\t$. We deal with the case of $\t=[c_1,c_2]$ for $0< c_1< c_2< 1$, and the other situations of $\t$ can be done by the similar way. In this case, for any distribution $\mu\in\ds$, $\mu\models\neg a$ if and only if $\mu\models a_1\lor a_2$, i.e. $\neg a=a_1\lor a_2$, where $a_1=\pair{k}{[0, c_1)}$ and $a_2=\pair{k}{(c_2,1]}$. Therefore,
$$\Phi=\land_{j\in J'}\lor_{ l \in L'_{j}}a_{j,l}$$ for some finite sets $J'$, $L'_{j}$ for $j\in J'$, $\{a_{j,l}\in \ap\}_{j\in J', l\in L'_j}$.

From $\mu_t\models \land_{j\in J'}\lor_{ l \in L'_{j}}a_{j,l}$, by the semantic of CLL, we have that for all $j\in J'$, $\mu_t\models a_{j,l}$ for some $l\in L'_j$. As $J'$ and all $L'_j$ are finite sets, we can check one by one whether or not $\mu_t\models a_{j,l}$. Let $\tt_{j,l}$ be the set of all the maximum intervals such that for each $\t\in\tt_{j,l}$, $\mu_t\models a_{j,l}$ for all $t\in \t$. Then $$\tt=\cap_{j\in J'}\tt_{j},$$
where $$\tt_{j}=\{\t_{1}\cup\t_{2}: \t_1\in \tt_{j,l_1} \textrm{ and } \t_2\in \tt_{j,l_2} \textrm{ for all distinct } l_1,l_2\in L'_j\} \textrm{ for all } j\in J',$$ and $$\tt_{i}\cap\tt_{j}=\{\t_1\cap\t_2: \t_1\in \tt_{i} \textrm{ and } \t_2\in \tt_{ j}\} \textrm{ for all } i,j\in J'.$$
The left problem is to handle $\t_1\cap\t_2$ and $\t_1\cup\t_2$, and by the continuity of PEFs, the left and right endpoints of $\i_1$ and $\i_2$ are all the roots of PEFs. It is equivalent to compare the values of two real roots of two (different) PEFs. This can be done by Lemma~\ref{lem:compare}.
\end{proof}

\subsection{Proof of Lemma~\ref{Lem:characterization}}
\begin{proof}
We first prove the sufficient direction. Let \[\vp_1=\Phi_1U^{\t_2}\Phi_2\cdots U^{\t_n}\Phi_{n}.\] Then the above formula is $\Phi_0U^{\t_1}\vp_1$. By the semantic of CLL, we have that there is a time $t_1\in\t_1$ such that $\sigma_{\mu_{t_1}} \models \vp_1$, and for any $t_1'\in [0,t)\cap\t_1$, ${\mu_{t_1'}}\models \Phi_0$. Then let \[\vp_2=\Phi_2U^{\t_3}\Phi_3\cdots U^{\t_n}\Phi_{n}\] and we get $\vp_{1}=\Phi_1U^{\t_2}\vp_2$. In the similar way, we have that there is a time $t_2\in\t_2$ such that $\sigma_{\mu_{t_1+t_2}} \models \vp_2$, and for any $t_2'\in[0 ,t_2)\cap \t_2$, ${\mu_{t_1+t_2'}}\models \Phi_1$. Iteratively, we get a set of time instants $\{t_{k}\}_{k=0}^{n}$ with $t_0=0$. For all $1\leq k\leq n$, let
\[\i_{k}=\sum_{j=0}^{k-1}t_{j}+[0,t_{k})\cap\t_k.\]
Then it is easy to check that $\{\i_{k}\in \r^{+}\}_{k=0}^{}$ with $\i_{0}=[0,0]$ are the desired intervals satisfying the above two conditions.

Considering the necessary direction, by the above proof, we only need to identify $\{t_k\}_{k=1}^{n}$ throughout intervals $\{\i_k\}_{k=0}^{n}$. Let $t_{k}=\sup\i_{k}-\sup\i_{k-1}$ for all $1\leq k\leq n$.

\end{proof}

\subsection{Proof of Theorem~\ref{Thm:isolating}}\label{appendix:theorem}
Before presenting the proof of Theorem~\ref{Thm:isolating}, we recall one useful technique from previous works --- factoring PEFs.

\subsubsection{Factoring PEFs}
An essential step of finding a real root isolation of a PEF is factoring it into a set of PEFs, such that the resulting PEFs have no multiple roots except for zero. In the following, we introduce a technique to implement this. Before that, we recall some concepts.

Given a root $t^*$ of a function $f(t)$, i.e., $f(t^*)=0$, the multiplicity of $t^*$ is the maximum number $m$ such that $(t-t^*)^m$ is a factor of $f(t)$, i.e., there exists a function $g(t)$ such that $f(t)=g(t)(t-t^*)^m$. In particular, if $m=1$, then we call that $t^*$ is a \emph{single root}; otherwise $t^*$ is a \emph{multiple root}.

Recall a PEF $f:\reals\to\reals$ has the following form:
\begin{equation} f(t)=\sum_{k=0}^{K}f_k(t)e^{\lambda_k t}
\end{equation}
where for all $0\leq k\leq K<\infty$, $f_k(t)\in\a[t]$, $\lambda_k\in\a$.
Without loss of generality, we assume $\lambda_k$'s are distinct, and use $\Exp(f)$ to denote the set of \emph{exponent} of $f(t)$, i.e., $\Exp(f)=\{\lambda_{k}\}_{k=0}^{K}$.

Let $\{a_{j}\}_{j=1}^n$ be a linearly independent basis over $\q$ of the exponents $\Exp(f)=\{\lambda_k\}_{k=0}^K$ appearing in $f(t)$ such that $f(t)$ is a multivariate polynomial with respect to $t, e^{a_1 t},\ldots, e^{a_nt}$, denoted by \[f(t, e^{a_1 t},\ldots, e^{a_nt}).\]
This basis can be obtained by
computing a simple extension (e.g. \cite{loos1982computing}, \cite[Algorithm~1]{huang2018positive})

$$\mathbb{Q}(\lambda)/\mathbb{Q}=\mathbb{Q}(\lambda_0,\ldots,\lambda_K)/\mathbb{Q}.$$
%In particular, let $P_0,\ldots,P_n$ be the minimal polynomials in Definition~\ref{def:algebraic_number} of $\lambda_0,\ldots,\lambda_{n}$, respectively. such that each $\lambda_i=P_i(\lambda)$ with $P_i\in \mathbb{Q}[x]$ and $\deg(P_i)<\deg(\lambda)$, where $\deg(P_i)$ is the degree of polynomial $P_i$, i.e., the highest of the degrees of the polynomial's monomials (individual terms) with non-zero coefficients. Furthermore, choosing a better basis $\{\gamma_1,\ldots,\gamma_L\}$,such that each $\lambda_i$ can be $\mathbb{Z}^{+}$-linearly expressed by $\{\gamma_1,\ldots,\gamma_L\}$.Thus we get a multivariate polynomial representation $g(t,e^{\gamma_1 t},\ldots,e^{\gamma_L t})$ of $f(t)$.
Next, we replace $t$ by the indeterminate $y_0$, and each exponential term $e^{a_i t}$ by $y_{i}$ for $1\leq i\leq n$.
Then, we have a multivariate polynomial representation $f(y_0,\ldots,y_n)$ of $f(t)$ in the $y's$. By the factorization of polynomials, we have
$$f(y_0,\ldots,y_n)=f_{1}^{m_1}(y_0,\ldots,y_n)\cdots f_{s}^{m_s}(y_0,\ldots,y_n)\quad \textit{for}\quad m_{1},\ldots, m_s\in\z^{+}$$
whose \emph{greatest square free part} is denoted by
\[\hat{f}(y_0,\ldots,y_n)=f_{1}(y_0,\ldots,y_n)\cdots f_{1}(y_0,\ldots,y_n).\]
Via switching $y's$ back by $e^{a_it}$ and $t$, we have a PEF
\begin{equation}\label{Eq:square_free}
\hat{f}(t)=\hat{f}(t, e^{a_1 t},\ldots, e^{a_nt}).
\end{equation} It is worth noting that the set of roots of $\hat{f}(t)$ is the same to the one of $f(t)$. In the latter, we will prove $\hat{f}(t)$ only has single roots.

At last, a corollary of Schanuel's conjecture is also needed to prove Theorem~\ref{Thm:isolating}.

\begin{corollary}\label{Cor:degree}\cite[Corollary 3]{gan2017reachability}
Let $ a_1 , \ldots, a_n$ be algebraic numbers that are
linearly independent over $\q$. Under the condition that Schanuel's conjecture holds, the transcendence degree of the extension field $\q(t,e^{a_1 t}, \ldots, e^{a_n t})$ is at least $n$ if $t \not= 0$.
\end{corollary}
\subsubsection{Main Proof}
In this subsection, we prove Theorem~\ref{Thm:isolating} with the help of the exclusion algorithm in~\cite[Algorithm 2]{huang2018positive} which can compute a real root isolation $\iso{(f)}_{\t}$ for a given simple PEF $f(t)$ and interval $\t$. Here, a PEF $f(t)$ is simple if all roots of $f(t)$ are single.

In particular, if $f(t)$ has only single roots in $\t$ (there is no $t\in \t$ such that $f(t)=f'(t)=0$), then we can get a real root isolation $\iso(f)_\t$ by \cite[Algorithm 2]{huang2018positive}.
Otherwise, $f(t)$ has at least one multiple root, and thus the exclusion algorithm in~\cite{huang2018positive} cannot be used to compute a real root isolation of $f(t)$ directly. This can be dealt with by employing factoring PEFs in the last section and the following fact.

\begin{lemma}\label{Lem:multiple_case}
Let $f(t)=f(t, e^{a_1 t},\ldots, e^{a_nt})$ be a PEF with respect to $t$, and thus a polynomial with respect to $t, e^{a_1 t},\ldots, e^{a_nt}$, where $\{a_i\}_{i=1}^n$ is linearly independent over $\q$. If the multivariate polynomial representation $f(y_0,\ldots,y_n)$ of $f(t)$ is square free, then,
assuming that Schanuel's conjecture holds, $f(t)$ has no multiple real root except $0$.
\end{lemma}
\begin{proof} %The proof is similar to \cite[Corollary 4]{gan2017reachability} (PEFs over real numbers). For the completeness, we present the proof as follows.
Since $f(y_0,\ldots,y_n)$ is square free, we may write
$$f(y_0,\ldots,y_n)=f_{1}(y_0,\ldots,y_n)\cdots f_{1}(y_0,\ldots,y_n)$$
where for any $1 \leq i,j \leq n$, $i \not= j$, $f_{i}(y_0,\ldots,y_n)$ is irreducible, and $f_{i}(y_0,\ldots,y_n)$ and $f_{j}(y_0,\ldots,y_n)$ are coprime, i.e., they do not share a polynomial $g(y_0,\ldots,y_n)$ as a factor.

By contradiction, we first prove that \[f_i (t,e^{a_1 t},\ldots, e^{a_n t})
\textit{ and } f_j(t,e^{a_1 t},\ldots,e^{a_n t})\]
have no nonzero common real root with respect to $t$.
Assume that $t_0 \not=0$ is a common real root of $f_i (t,e^{a_1 t},\ldots, e^{a_n t})$
and $f_j(t,e^{a_1 t},\ldots,e^{a_n t})$. By Corollary~\ref{Cor:degree}, we have that the
transcendence degree of $\q(t_0,e^{a_1 t_0} ,\ldots,e^{a_n t_0} )$ is at least $n$.
Thus there must exist $n$ elements in $\{t_0,e^{a_1 t_0} ,\ldots,e^{a_n t_0}\}$
that are algebraically independent. Without loss of generality, let $\{t_0,e^{a_1 t_0} ,\ldots,e^{a_{n-1} t_0}\}$ be the $n$ elements that are algebraically independent. Considering the multivariate polynomial representation, let $g(y_0,\ldots,y_n)$ be the
resultant~\cite[Page 26]{trott2007mathematica} of $f_{i}(y_0,\ldots,y_n)$ and $f_{j}(y_0,\ldots,y_n)$
with respect to $y_n$.
Then $(t_0,e^{a_1 t_0} ,\ldots,e^{a_{n-1} t_0})$ is a root of $g(y_0,y_1,\ldots,y_{n-1})$, which is a nontrivial polynomial
as \[f_i (t,e^{a_1 t},\ldots, e^{a_n t})
\textit{ and } f_j(t,e^{a_1 t},\ldots,e^{a_n t})\] are coprime. Therefore, $(t_0,e^{a_1 t_0},\ldots,e^{a_n t_0})$ is a root of some nontrivial
polynomial. This contradicts that $\{t_0,e^{a_1 t_0},\ldots,e^{a_n t_0}\}$ are
algebraically independent.

Next, we prove that $f_i(t,e^{a_1 t},\ldots, e^{a_n t})$ has no multiple real
root. Suppose
$$f_i(t,e^{a_1 t},\ldots, e^{a_n t})=h_0(t)+\sum_{k=1}^s h_k(t)(e^{a_1t})^{b_{k1}} \cdots(e^{a_n t})^{b_{kn}}$$
where $h_0 (t), \ldots , h_n (t)$ are nontrivial polynomials and $b_{jk}\in\n$ for $1 \leq j \leq s$ and $1 \leq k \leq n$. Then we have
\begin{align*}
    &f_{i}'(t,e^{a_1t},\ldots, e^{a_n t})\\
    =&h_0'(t) + \sum_{k=1}^s[h_k'(t)+h_k'(t)(a_1b_{k1}+\cdots+a_nb_{kn})](e^{a_1t})^{b_{k1}} \cdots(e^{a_n t})^{b_{kn}}.
\end{align*}

Moreover, considering the corresponding multivariate polynomial representation,

\begin{align*}
    &f_i(y_0,y_1,\ldots, y_n)=h_0(y_0)+\sum_{k=1}^s h_k(y_0)(y_1)^{b_{k1}} \cdots(y_n)^{b_{kn}}\\
    &f_{i}^{'}(y_0,y_1,\ldots, y_n)\\
    =& h_0'(y_0) + \sum_{k=1}^s[h_k^{'}(y_0)+h_k^{'}(y_0)(a_1b_{k1}+\cdots+a_nb_{kn})](y_1)^{b_{k1}} \cdots(y_n)^{b_{kn}}.
\end{align*}
From the degree of $h_0(y_0)$ in the above two polynomials, it is evident to see that
$f_i(y_0,y_1,\ldots, y_n)$ is not a factor of $f_{i}^{'}(y_0,y_1,\ldots, y_n) $. Then, $f_i(y_0,y_1,\ldots, y_n)$ and $f_{i}^{'}(y_0,y_1,\ldots, y_n)$ are coprime, since $f_i(y_0,y_1,\ldots, y_n)$ is irreducible. For the same reason as above, $f_i(t,e^{a_1 t},\ldots, e^{a_n t})$ and $f_{i}^{'}(t,e^{a_1t},\ldots, e^{a_n t})$ have no common real roots. Therefore, $f_i(t,e^{a_1 t},\ldots, e^{a_n t})$ has no multiple real root.
\end{proof}%For the readability, the proof of the above lemma will be delayed to the next section.

By the above lemma, for finding a real root isolation $\iso{(f)}_\t$ of a PEF $f(t)$ in some time interval $\t$, we first implement the method of factoring PEFs in the last section to get a simple PEF $\hat{f}(t)$, which shares the same roots with $f(t)$. Then we can implement the exclusion algorithm in~\cite{huang2018positive} to achieve this goal. Finally, we complete the proof of Theorem~\ref{Thm:isolating} by noting that the number of obtained real root isolations by the algorithm is finite (see the correctness analysis of \cite[Algorithm 2]{huang2018positive}), i.e., $|\iso{(f)}_\t|<\infty$.

\end{document}